\documentclass[reprint,aps]{revtex4-1}
\usepackage{amssymb,amscd,amsmath,amsthm}
\usepackage{latexsym,amstext}
\usepackage{dcolumn}
\usepackage{color}

\def\be{\begin{equation}}
\def\ee{\end{equation}}
\def\bea{\begin{eqnarray}}
\def\eea{\end{eqnarray}}
\def\bean{\begin{eqnarray*}}
\def\eean{\end{eqnarray*}}
\def\ds{\displaystyle}
\def\be{\begin{equation}}
\def\ee{\end{equation}}

\def\N{{\mathbb N}}
\def\Z{{\mathbb Z}}

\def\C{{\mathbb C}}
\def\ca{{\mathcal C}}

\def\H{{\mathcal H }}

\def\O{\Omega}

\def\l{\langle}
\def\r{\rangle}

\begin{document}

\title{Spherical harmonics and rigged Hilbert spaces}

\author{E. Celeghini}
 \email{celeghini@fi.infn.it}
 \affiliation{Dipartimento di Fisica, Universit\`a di Firenze and\\  INFN-Sezione di Firenze
150019
Sesto Fiorentino, Firenze, Italy.}

\author{M. Gadella}
 \email{manuelgadella1@gmail.com}

 \author{M. A. del Olmo}
 \email{marianoantonio.olmo@uva.es}
\affiliation{Departamento de F\'{\i}sica Te\'orica, At\'omica y \'Optica and IMUVA. \\
Universidad de Valladolid, Paseo Bel\'en 7, 47011 Valladolid, Spain.
}

\date{\today}

%%%%%%%%%%%%%%%%%%
\begin{abstract}
This paper is devoted to study discrete and continuous bases for spaces supporting representations of $SO(3)$ and $SO(3,2)$
where the  spherical harmonics are involved. We show how discrete and continuous bases coexist on appropriate choices of rigged Hilbert spaces. We prove the continuity of relevant operators and the 
operators in the algebras spanned by them using appropriate topologies on our spaces. Finally, we discuss the properties of the functionals that form the continuous basis. 
\begin{description}
\item[PACS numbers]
02.20.Sv, 02.30.Gp, 02.30.Nw, 03.65.Ca
\end{description}
\end{abstract}

%Keywords: quantum mechanics, signal theory, Hilbert spaces, spherical harmonics

\maketitle
%%%%%%%%%%%%%%%%%%%%%%%%%%%%%%%%%%%%%%%%%%%%%%
\section{Introduction}
%%%%%%%%%%%%%%%%%%%%%%%%%%%%%%%

As is well known, spherical harmonics (SH) are of wide use in applied as well as in theoretical physics, and more recently in signal processing. In reference to this latter
 field, applications  include  the analysis of  signals on the sphere in geodesy, astronomy and cosmology, like the study of cosmic microwave background.   SH are  also considered for 3D modelling  in graph computation, vision computation,  medical image and communication systems \cite{KS,RH,MRC}. 

Closely related to SH are the associated Legendre functions, which are used  to construct support spaces for representations of the Lie algebras $so(3)$ and $so(3,2)$.  Also, SH   support  the same representations \cite{CO}.   An interesting feature   is that both kinds of special functions  somehow serve as discrete and continuous basis for their representation space. The discrete basis is the orthonormal basis labelled by the indices $l=0,1,\dots$ and $m=-l,\dots,l$. . In the case of SH, the continuous basis is labeled by the values of the angles $\theta$ an $\phi$, but only the  angle $\theta$ labels the continuous basis for the associated Legendre polynomials.  We shall go into the details along the present manuscript. 

In a previous paper \cite{CGO}, we have provided   both discrete and continuous bases labelled by the whole and half real line, again in connection to Lie algebras and signal theory. There, we have used support spaces constructed with the use of Hermite and Laguerre functions, respectively.  
Since the supporting spaces are in any case separable, the only manner to introduce both basis in the same structure is to equip the supporting Hilbert space so as to have a rigged Hilbert space (RHS), also called Gelfand triplet. All three spaces in the RHS support the representation and the elements of the enveloping algebras are continuous on the spaces other than the Hilbert space.

Motivated by our results in \cite{CGO}, we plan to extend the formalism to other situations in which representations of Lie algebras naturally produce the need for both discrete and continuous basis. In \cite{AP}, we have studied the case for $SO(2)$. In the present paper, we extend the formalism for the mentioned cases of $SO(3)$ and  $SO(3,2)$ with quadratic Casimir $\mathcal C_{so(3,2)}=-5/4$. Further examples will be studied on later works. 

Also in physics there are several  and independent reasons to assert that Hilbert spaces are not sufficient for a thorough formulation of Quantum Mechanics even within the non-relativistic context. To begin with the Dirac formulation \cite{D,BO,RO,ANT,MELSH,GG,GG1}, where eigenvectors of 
operators with eigenvalues in the 
continuous spectrum play a crucial role. Furthermore, these eigenvectors, not in the Hilbert space of square integrable wave functions, are widely used in quantum mechanics. A proper definition of Gamow vectors, which are commonly used in calculations including unstable quantum systems \cite{KKH,BO1,BG,CG}, are also non-normalizable. Finallly, formulations of time asymmetry in quantum mechanics requires the use of tools more general  than Hilbert spaces, but including them,  like rigged Hilbert spaces constructed with Hardy functions \cite{BO2,BGK,MAR}.

Since RHS are suitable frameworks where   discrete and continuous bases are involved, 
 a typical situation for special functions where   discrete  labels and   continuous variables appear, 
it looks reasonable to conjecture that there exists a  close relation between RHS, discrete and continuous bases and special functions, where the existence of one of these ingredients implies the need for the others.

To begin with, in Section 2 we give a brief presentation of rigged Hilbert spaces and some relevant properties of interest. In Section 3, we recall some properties of SH along some considerations about discrete and continuous bases and their relations. Section 4 contains the main results of this presentation. It is shown the continuity on a given rigged Hilbert space of the relevant operators related with $SO(3,2)$. All they are unbounded operators from the point of view of Hilbert space and belong to two different categories: i.) either are unbounded generators   of the algebra; ii.) or they are unbounded relevant functions of the bounded generators of the algebra. We close the section with a discussion on the mathematical meaning of the continuous basis. We end this paper with some concluding remarks.

%%%%%%%%%%%%%%%%%%%%%%%%%%%%%%%%%%%%%%%%%%%%%%%%
%%%%%%%%%%%%%%%%%%%%%%%%%%%%%%%%%%%%%%%%%%%%%%%%
\section{Rigged Hilbert spaces: an overview}\label{RHSoverview}

Although RHS have been widely used in scientific publications not all readers may be familiar with this notion. It is the purpose of the present Section to define the concept of RHS and mention some of their most relevant properties. A RHS is a triplet of spaces \cite{GEL}:

\begin{equation}\label{1}
\Phi\subset\mathcal H\subset \Phi^\times\,,
\end{equation}
where $\mathcal H$ is an infinite dimensional (separable) Hilbert space. The {\it space of test vectors}  $\Phi$ is a dense subspace of $\mathcal H$ endowed with its own topology, which is finer or stronger (contains more open sets) than the topology  of $\mathcal H$. 
A straightforward consequence of this fact is that all sequences which converge on $\Phi$, also converge on $\mathcal H$, the converse being not true. The difference between topologies on $\Phi$ and $\mathcal H$ gives rise  that the dual space of $\Phi$, $\Phi^\times$,  is bigger than $\mathcal H$, which is self-dual. 

In this paper we shall consider the anti-dual $\Phi^\times$ of $\Phi$, instead the dual. This means that any $F\in\Phi^\times$ is a continuous anti-linear mapping from $\Phi$ into $\C$, where anti-linearity means that   for all $ \psi,\varphi\in\Phi$ and for all  $\alpha,\beta\in \C\,,$ we have
\begin{equation}\label{2}
\langle \alpha\psi+\beta\varphi|F\rangle=\alpha^*\langle\psi|F\rangle
+\beta^*\langle\varphi|F\rangle\,,
\end{equation}
where the star denotes complex conjugation. Here, we have represented the action of an arbitrary $F\in\Phi^\times$ on an arbitrary  $\psi\in\Phi$ as $\langle\psi|F\rangle$, in order to accommodate to the Dirac bra-ket notation. For the same reasons, we use anti-linear instead of linear mappings.  The antidual space $\Phi^\times$ may be endowed with any topology compatible with the dual pair $(\Phi,\Phi^\times)$, in particular with the weak topology. 

Another important property to be mentioned is the following one: let $A$ be a densely defined operator on $\mathcal H$ such that $\Phi$ be a subspace of its domain and $A\varphi\in\Phi\,,\;\forall  \varphi\in\Phi$. Then, we say that $\Phi$ reduces $A$ or, equivalently, that $\Phi$ is invariant under the action of $A$ i.e.,  $A\Phi\subset\Phi$.  In this case, $A$ may be extended unambiguously to the anti dual $\Phi^\times$ by making use of the duality formula:
\begin{equation}\label{3}
\langle A\varphi|F\rangle=\langle \varphi|A^\times F\rangle\,,\quad \forall\,\varphi\in\Phi\,, 
\;\forall\,F\in\Phi^\times\,.
\end{equation}
If, in addition, $A$ is continuous on $\Phi$,  its extension to $\Phi^\times$, $A^\times$ , is continuous on $\Phi^\times$ and, hereafter,  is also denoted by $A$   for simplicity.

Although this is not the most general case, in the examples we deal with in the present work, the topology on $\Phi$ is given by an infinite countable set of norms, say $\{ \big|\big|- \big|\big|_{n=1}^\infty\}$. Then, a linear operator $A$ on $\Phi$ is continuous if and only if for each norm $ \big|\big|- \big|\big|_n$ there is a positive number $K_n$ and a finite sequence of norms 
$\big|\big|-\big|\big|_{p_1},  \big|\big|- \big|\big|_{p_2}, \dots,  \big|\big|- \big|\big|_{p_r}$ such that for any $\varphi\in\Phi$, one has \cite{RS}:
\begin{equation}\label{4}
 ||A\varphi ||_n\le K_n\,\sum_{i=1}^r ||\varphi ||_{p_i}
\end{equation}
The same result applies to check the continuity of any linear or antilinear mapping $F:\Phi\longmapsto\mathbb C$, where $\mathbb C$ is the field of complex numbers.  In this case, the norm $ \big|\big|A\varphi \big|\big|_n$ in (\ref{4}) should be replaced by the modulus $|F(\varphi)|$. 
%%%%%%%%%%%%%%%%%%%%%%%%%%%%%%%%%%%%%%
%%%%%%%%%%%%%%%%%%%%%%%%%%%%%%%%%%%%%%
\section{Spherical harmonics}

In a previous paper \cite{AP}, we have given a  RHS   supporting  unitary irreducible representations  for $SO(2)$. This is a quite simple example, in which the simultaneous use of continuous and discrete bases, as is customary in the standard quantum mechanics, may be used without giving up mathematical rigor. At the same time, the elements of the Lie algebra $so(2)$ and of its corresponding enveloping algebra are continuous on the spaces $\Phi$ and $\Phi^\times$ that equip the Hilbert space. Other cases are studied in \cite{CGO}. 

As announced in the Introduction, we here study a different situation in which we use special functions to generate the RHS. Instead of Hermite or Laguerre functions, we here use SH.  They are functions of two angular variables, $\theta$ and $\phi$, and are defined on the hull $S^2$ of the 3-D unit sphere. It is well known that for $l$ fixed SH span a Hilbert space supporting a UIR of $SO(3)$, although it is less acknowledged that for $l\in \N$  they span a space supporting a UIR with Casimir $C_{so(3,2)}=-5/4$ of the de Sitter group $SO(3,2)$ \cite{CO}, as already mentioned.  Properties of SH are well known, so that we refer the reader to the standard literature on the subject \cite{CT,AS}. In the next subsection, we recall some important facts with a presentation which will help with our discussion.  

%%%%%%%%%%%%%%%%
%%%%%%%%%%%%%%%%
\subsection{Spherical harmonics: an overview}

Let us consider the hollow unit sphere $S^2$ in $\mathbb R^3$. Any point of $S^2$ is characterized by two angular variables $\theta$ and $\phi$, whose range of variation is $0\le \theta\le \pi$ and $0\le \phi< 2\pi$.  We define the spherical harmonics as \cite{CO}:
\begin{equation}\label{5}
Y_l^m(\theta,\phi) = \sqrt{\frac{(l-m)!}{2 \pi(l+m)!}}\;  e^{i m \phi} \;  P_l^m(\cos \theta)\,,
\end{equation}
in such a way that they satisfy eq.~\eqref{41} that is the standard form of the algebra $so(3,2)$  (see 
\cite{CO} for more details),  $l\in\mathbb N$, $m\in\mathbb Z$ with $|m|\le l$ and $P_l^m$ are the associated Legendre functions.    SH verify the following differential equation:
\begin{widetext}
\begin{equation}\label{6}
\left(\frac{1}{\sin\theta}\,\frac{\partial}{\partial \theta}\;\sin\theta\;{\partial \theta}+\frac{1}{\sin^2\theta}\,
\frac{\partial^2}{\partial \phi^2} +l(l+1)\right)\;Y_l^m(\theta,\phi) =0 \,.
\end{equation}
\end{widetext}
SH are the support space of a UIR of $SO(3,2)$ with quadratic Casimir ${\mathcal C}_{so(3,2)}=-5/4$. 
The action of the operators of the Cartan subalgebra of the Lie algebra  $so(3,2)$, (it  has rank 2 since it is a noncompact real form of $B_2$), $L$ and $M$, on SH  is
\be\begin{array}{lll}\label{06}
L\,Y_l^m(\theta,\phi)&=&l\;Y_l^m(\theta,\phi)\,,\\[0.3cm]
M\,Y_l^m(\theta,\phi)&=&m\;Y_l^m(\theta,\phi)\,.
\end{array}\ee
 SH $\{Y_l^m(\theta,\phi)\}$ form an orthonormal basis for the Hilbert space $L^2(S^2,d\Omega)$ of Lebesgue square integrable functions on $S^2$ with measure $d\Omega:= d(\cos\theta)\,d\phi$. Any function $f(\theta,\phi)\in L^2(S^2,d\Omega)$ admits the span
\begin{equation}\label{7}
f(\theta,\phi)=\sum_{l=0}^\infty \sum_{m=-l}^l  f_{l,m}\,\sqrt{l+1/2}\,Y_l^m(\theta,\phi)\,,
\end{equation}
where the series converges in the sense of the norm in $L^2(S^2,d\Omega)$. As is well known, a necessary and sufficient condition for it is that 
\begin{equation}\label{8}
\sum_{l=0}^\infty \sum_{m=-l}^l | f_{l,m}|^2<\infty\,.
\end{equation}
Coefficients $f_{l,m}$ in (\ref{7}) are given by the expression:
\[\label{9}
 f_{l,m}= \int_0^{2\pi} d\phi \int_0^\pi d(\cos\theta)\,Y_l^m(\theta,\phi)^*\,f(\theta,\phi)\,,
\]
where the star denotes complex conjugation. 

Since SH form an orthonormal basis, they  satisfy   relations of orthonormality and completeness  \cite{CO}:
\begin{widetext}\begin{equation}\label{10}
\begin{array}{rll}\ds
\int d\Omega\;\, Y_l^{m}\;(\theta,\phi)^*\;(l+1/2)\; Y_{l'}^{m'}(\theta,\phi) &=& \delta_{l,l'}\, \delta_{m,m'}\,,
\\[0.4cm]\ds
\sum_{l=|m|}^{+\infty}\sum_{m=-\infty}^{+\infty} Y_l^{m}\;(\theta,\phi)^*\;(l+1/2)\;   Y_{l}^{m}(\theta',\phi') &=& \delta\left(\cos\theta-\cos\theta'\right)\,   \delta(\phi-\phi')\,,
\end{array}\end{equation}
\end{widetext}
with $\delta\left(\cos\theta-\cos\theta'\right)=\delta(\theta-\theta')/\sin\theta$, where the presence of tempered distributions and Kronecker deltas is a signal that rigged Hilbert spaces must be involved.  

We also consider an abstract Hilbert space $\H$ support of an equivalent UIR of $SO(3,2)$ of Casimir $\ca_{so(3,2)}=-5/4$, mentioned above. Let $\widetilde L,\widetilde M$ the elements of the Cartan subalgebra of $so(3,2)$ in this representation so that
\begin{equation}\label{11}
\widetilde L\,|l,m\rangle=l\,|l,m\rangle\,,\quad \widetilde M\,|l,m\rangle=m\,|l,m\rangle\;,
\end{equation}
where $ l=0,1,2,\dots$ and $-l\leq m\leq l\in\Z\,.$
The set of eigenvectors  $\{|l,m\rangle\}$ is a basis for $\H$ fulfilling 
the properties of orthogonality and completeness
\begin{equation}\begin{array}{l}\label{12}
\langle l,m|l',m'\rangle=\delta_{l,l'}\,\delta_{m,m'}\,, \\[0.3cm] 
 \ds\sum_{l=0}^\infty \sum_{m=-l}^l |l,m\rangle\langle l,m|=\mathcal I\,,
\end{array}\end{equation}
where $\mathcal I$ is the identity on $\H$. Any $|f\r\in\mathcal H$ may be written as
\begin{equation}\label{13}
|f\r=\sum_{l=0}^\infty\sum_{m=-l}^l f_{l,m}\,|l,m\rangle\,,
\end{equation}
with $f_{l,m}=\langle l,m|f\rangle$, if and only if 
\begin{equation}\label{14}
 \sum_{l=0}^\infty\sum_{m=-l}^l \big|f_{l,m}\big|^2<\infty\,.
\end{equation}

We may easily establish a canonical bijection of $\mathcal H$ on $L^2(S^2,d\Omega)$ by
 \begin{equation}\label{15}
 \begin{array}{lccc}
 S:&\mathcal H&\longmapsto &L^2(S^2,d\Omega)\\[0.4cm]
 &|l,m\rangle&\longmapsto&\;\; \sqrt{l+1/2}\; Y^m_l (\theta,\phi) \,,
\end{array}
\end{equation}
and extended by linearity and continuity to the whole $\mathcal H$. It is simple to check that $ S$ is a unitary mapping. Note that we have to use relations \eqref{10}. Since $S\,|l,m\rangle=\sqrt{l+1/2}\;Y^m_l (\theta,\phi)$ we have obviously that 
$S^{-1}\,\left(Y^m_l (\theta,\phi)\right)=|l,m\rangle/\sqrt{l+1/2}$ and we  relate the operators $L,M$ \eqref{06} and $\widetilde L,\widetilde M$ \eqref{11} by  $\widetilde L=S^{-1}\,L\,S$ and $\widetilde M=S^{-1}\,M\,S$.

For any $|f\r \in\H$ satisfying \eqref{13}, we have the following expression:
\begin{equation}\begin{array}{lll}\label{16}
S\,|f\r =\ds\sum_{l=0}^\infty\sum_{m=-l}^l f_{l,m}\,S\,|l,m\rangle\\[0.3cm]
\qquad=\ds\sum_{l=0}^\infty\sum_{m=-l}^l f_{l,m}\,\sqrt{l+1/2}\, Y^m_l (\theta,\phi)\,.
\end{array}\end{equation}

We  now introduce a continuous basis, $\{|\theta,\phi\rangle\}$, depending on the values of the angles $\theta$ and $\phi$ with the help of the discrete basis $\{|l,m\r\}$ by
\begin{equation}\label{17}
\langle \theta,\phi |l,m\rangle :=\sqrt{l+1/2}\,Y_l^m(\theta,\phi)\,,
\end{equation}
expression which is defined for (almost) all fixed values of $\theta$ and $\phi$, with $0\le\theta\le\pi$ and $0\le\phi<0$. 
For almost all fixed values of  $\theta$ and $\phi$, equation \eqref{17} may be extended to any  
$|f\r \in\mathfrak G$, where $\mathfrak G$ is a dense subspace in $\mathcal H$ with properties to be determined later, making use of \eqref{16}. The result is
\begin{equation}\begin{array}{l}\label{18}
\langle\theta,\phi|f\rangle= f(\theta,\phi):= S\, |f\r\\[0.3cm]
\quad=\ds\sum_{l=0}^\infty\sum_{m=-l}^l f_{l,m}\, \sqrt{l+1/2}\,Y_l^m(\theta,\phi)\,.
\end{array}\end{equation}
We recoover the expansion \eqref{7}. Relations \eqref{18} should be looked as functions on the variables $\theta$ and $\phi$ and the series converges in the sense of the norm on $L^2(S^2,d\Omega)$. For almost all fixed values of $\theta$ and $\phi$, the relation $\langle\theta,\phi|f\rangle= f(\theta,\phi)$ is also valid. 

It is important to remark that $\langle\theta,\phi|$ cannot be defined as a continuous mapping on the Hilbert space $\mathcal H$, although quantum mechanics textbooks give the first identity in \eqref{18}, which for any $|f\r\in\mathcal H$ is indeed merely formal. As a matter of fact, we shall define in a proper way $\langle\theta,\phi|$ as a linear continuous functional over a space $\mathfrak G$, with the property that the triplet $\mathfrak G\subset\mathcal H \subset \mathfrak G^\times$ be a RHS. Here, $|\theta,\phi\rangle\in\mathfrak G^\times$, as we shall discuss later. Note that $\langle f|\theta,\phi\rangle:=\langle \theta,\phi|f\rangle^*$, where the star denotes complex conjugation. 

Next, we give some standard formulas that will have proper meaning when we define the space $\mathfrak G$ in the next section, including the meaning of the identity $\langle\theta,\phi|f\rangle=f(\theta,\phi)=S\,|f\r$, when $f(\theta,\phi)$ is considered as a function on its arguments on their  range. Thus, for any $|f\r,|g\r\in \mathfrak G$ and taking into account that $S\,|f\r,S\,|g\r\in L^2(S^2,d\Omega)$ with $S$ unitary, we have
\begin{equation}\begin{array}{l}\label{19}
\l f|g\r =\ds\l S f|Sg\r=\int_{S^2}\,d\Omega\, f^*(\theta,\phi)\,g(\theta,\phi)\\[0.3cm]
\qquad=\ds\int_{S^2}\,d\Omega \, \langle  f|\theta,\phi \r\,\langle  \theta,\phi |g\r\,.
\end{array}\end{equation}
From \eqref{19},   we may derive the completeness relation given by
\begin{equation}\begin{array}{lll}\label{20} 
\mathbb I &=&\ds \int_{S^2} d\Omega\, |\theta,\phi\rangle \langle \theta,\phi|\\[0,3cm]
&=&\ds \int_0^{2\pi}d\phi\int_0^{\pi}\;d(\cos\theta)\,|\theta,\phi\rangle \langle \theta,\phi|\,,
\end{array}\end{equation}
where $\mathbb I$ is a formal identity whose exact meaning will be discussed later.  Applying $\mathbb I$ on $|f\r\in \mathfrak G$ we find a well known formal formula for $\mathbb I$:

\begin{equation}\label{21}
{\mathbb I}\,|f\r=|f\rangle=\int_{S^2} d\Omega\, \l\theta,\phi |f\r\,|\theta,\phi\r\,.
\end{equation}
A standard formula may be obtained with the use of \eqref{20}. From
\begin{equation}\begin{array}{lll}\label{22}
f(\theta,\phi)&=&\ds \langle\theta,\phi|f\rangle\\[0.3cm]
&=&\ds \int_{S^2} d\O\,\langle\theta,\phi| \theta',\phi'\r \, \l\theta',\phi' |f\r\,,
\end{array}\end{equation}
we may derive the well known orthogonality relation for the kets $|\theta,\phi\r$ given by
\begin{equation}
\label{23}
\langle \theta,\phi\,|\,\theta',\phi'\rangle = \delta\left(\cos\theta-\cos\theta'\right)\, \delta(\phi-\phi')\,.
\end{equation}
Equation \eqref{20} shows that $|\theta,\phi\rangle$, for all values $0\le\theta\le\pi$ and $0\le\phi<2\pi$ plays the role of a generalized basis. We shall see the proper meaning of this assertion soon. 

We shall construct $\mathfrak G$ in such a way that $|l,m\rangle \in\mathfrak G$. Then,  go back to \eqref{21}, replace $|f\r$ by $|l,m\r$ and use \eqref{17}. This gives an expression of the {\it discrete} basis $\{|l,m\rangle\}$ in terms of the {\it continuous} basis $\{|\theta,\phi\rangle|\}$ given by
\[\label{24}
|l,m\rangle= \int_0^{2\pi} d\phi \int_0^\pi d\theta\,\sqrt{l+1/2}\,Y_l^m(\theta,\phi)\,|\theta,\phi\rangle\,.
\]
Observe that if we insert the identity $\mathcal I$, defined in (\ref{12}), in the l.h.s. of (\ref{18}), i.e. if we use the identity $\langle\theta,\phi|f\rangle=\langle \theta,\phi|\mathcal I|f\rangle$, we have
\begin{equation}\begin{array}{lll}\label{25}
f(\theta,\phi)&=&\ds\langle\theta,\phi|f\rangle=\langle \theta,\phi|\mathcal I|f\rangle\\[0.3cm]
&=&\ds\sum_{l=0}^\infty\sum_{m=-l}^l \langle \theta,\phi|l,m\rangle\langle l,m|f\rangle\,,
\end{array}\end{equation}
which shows that one may write $S$  as
\begin{equation}\begin{array}{lll}\label{26}
S&=&\ds\sum_{l=0}^\infty\sum_{m=-l}^l \langle \theta,\phi|l,m\rangle\langle l,m|\\[0.3cm]
&=&\ds \sum_{l=0}^\infty\sum_{m=-l}^l \sqrt{l+1/2}\,
Y_l^m(\theta,\phi)\,\langle l,m|\,.
\end{array}\end{equation}
From \eqref{21}, we see that
\begin{equation}\label{27}
|f\r= \int_{S^2} d\Omega\, f(\theta,\phi)\,|\theta,\phi\rangle\,,
\end{equation} 
which shows that for given $\theta$ and $\phi$ almost elsewhere with respect to the Lebesgue measure, $f(\theta,\phi)$ is the coefficient of $|\theta,\phi\rangle$ in the span of $|f\r\in\mathfrak G$ in terms of the continuous basis $\{|\theta,\phi\rangle\}$. 

As a general remark, note that $\mathbb I$ \eqref{20} has been constructed using the continuous basis, while $\mathcal I$ \eqref{12} is given by the discrete basis only. Furthermore, while  $\mathcal I$ is the identity on the Hilbert space $\mathcal H$, the  precise meaning of $\mathbb I$ will be clarified later. Taking an arbitrary pair of vectors $|f\r,|g\r\in\mathfrak G$ and using (\ref{17}), (\ref{25}) and (\ref{27}) we straightforwardly obtain the following two identities:
\begin{equation}\begin{array}{lll}\label{28}
\langle f|g\rangle&=&\ds\int_{S^2} \, d\O\,f^*(\theta,\phi)\,g(\theta,\phi)\\[0.3cm]
&=& \ds\sum_{l=0}^\infty\sum_{m=-l}^l f^*_{l,m} \,g_{l,m}
\end{array}\end{equation}
and
\begin{equation}\begin{array}{lll}\label{29}
 \big|\big|f \big|\big|^2&=&\ds \int_0^{2\pi} d\phi \int_0^\pi d(\cos\theta) \,|f(\theta,\phi)|^2\\[0.3cm]
 &=&\ds \sum_{l=0}^\infty\sum_{m=-l}^l  \big|f_{l,m}\big|^2\,.
\end{array}\end{equation}

This section has been devoted to a presentation on continuous and discrete bases in a form in use in the standard quantum mechanics.   As discrete basis, we have used a complete orthonormal set, $\{|l,m\rangle\}$ of eigenvectors of the Cartan operators $\widetilde L$ and $\widetilde M$. This opens the door for a discussion on other relevant operators acting on $\mathcal H$ or, equivalently through the unitary mapping $S$, on $L^2(S^2,d\Omega)$ that we will introduce in the next subsection.

%%%%%%%%%%%%%%%%%%%%%%%%%%%%%%%
%%%%%%%%%%%%%%%%%%%%%%%%%%%%%%%

\subsection{Meaningful operators for spherical harmonics.}

Let us consider an operator $\widetilde O$ densely defined on $\mathcal H$, bounded or not. 
Taking into account the unitary mapping $S :\mathcal H\longmapsto L^2(S^2,d\Omega)$, defined in the previous section \eqref{15}, the operator $O:=S\,\widetilde O\, S^{-1}$ acts on $L^2(S^2,d\Omega)$ and shares the properties with $\widetilde O$. Since $S\,|f\r=\langle\theta,\phi|f\rangle$, we have that

\begin{equation}\label{30}
 S\,\widetilde O |l,m\rangle=\langle\theta,\phi|\widetilde O|l,m\rangle\,.
 \end{equation}
 Using (\ref{26}), we obtain
\begin{equation}\begin{array}{lll}\label{31}
S\,\widetilde O|l,m\rangle &=&\ds S\,\widetilde O \, S^{-1}\, S|l,m\rangle\\[0.3cm]
&=&\ds\sqrt{l+1/2}\; O\,Y_l^m(\theta,\phi)\,,
\end{array}\end{equation}
provided that $|l,m\rangle$ lies in the domain of $\widetilde O$. Hence, we finally get
\begin{equation}\label{311}
O\,Y_l^m(\theta,\phi)=\ds\frac{1}{\sqrt{l+1/2}}\,\langle\theta,\phi |\widetilde O|l,m\rangle\,.
\end{equation}

Next, we introduce the operators which are under interest considered as operators on  $L^2(S^2,d\Omega)$. We know that they uniquely determine corresponding operators on $\mathcal H$ via $S$. These are:

\begin{enumerate}
\item
$\Theta$ and $\Phi$ are the multiplication operators by $\theta$ and $\phi$, respectively. As $\theta$ and $\phi$ are angular variables, both operators are bounded and, therefore, defined on the whole $L^2(S^2,d\Omega)$. 

\item
 The operators multiplication  by $l$ and $m$, respectively, $L$ and $M$, are unbounded.
 
 \item The operators   $D_\Theta:=d/d\theta$ and $D_\Phi:=d/d\phi$ are also unbounded. 
 \end{enumerate} 
 
 Following the above notation, we denote the corresponding operators on $\mathcal H$ as $\widetilde \Theta$, $\widetilde\Phi$, $\widetilde L$, $\widetilde M$, $\widetilde D_\Theta$ and $\widetilde D_\Phi$, respectively.  They act on the spherical armonic basis as follows \eqref{311}:
\begin{widetext}
\begin{equation}\begin{array}{lllll}\label{32}
\Theta\,Y_{l,m}(\theta,\phi)&=&\ds\frac{1}{\sqrt{l+1/2}}\,\langle \theta,\phi|\widetilde \Theta |l,m\rangle &:=& \theta \, Y_{l,m}(\theta,\phi)\,,
\\[0.35cm] 
\Phi\,Y_{l,m}(\theta,\phi) &=&\ds\frac{1}{\sqrt{l+1/2}}\,\langle \theta,\phi|\widetilde \Phi|l,m\rangle &:=&
 \phi \, Y_{l,m}(\theta,\phi)\,,
\\[0.35cm] 
L\,Y_{l,m}(\theta,\phi)&=&\ds\frac{1}{\sqrt{l+1/2}}\,\langle \theta,\phi|\widetilde L|l,m\rangle &:=& l\,Y_{l,m}(\theta,\phi)\,,
\\[0.35cm] 
M\,Y_{l,m}(\theta,\phi) &=&\ds\frac{1}{\sqrt{l+1/2}}\,\langle \theta,\phi|\widetilde M|l,m\rangle&:=&m\,Y_{l,m}(\theta,\phi)\,,
\\[0.35cm] 
D_\Theta\,Y_{l,m}(\theta,\phi)&=&\ds\frac{1}{\sqrt{l+1/2}}\,\langle \theta,\phi|\widetilde D_\Theta|l,m\rangle &:=&\ds\frac{\partial}{\partial\theta} 
\, Y_{l,m}(\theta,\phi)\,,
\\[0.35cm] 
D_\Phi\,Y_{l,m}(\theta,\phi)&=&\ds\frac{1}{\sqrt{l+1/2}}\,\langle \theta,\phi|\widetilde D_\Phi|l,m\rangle
&:=&\ds \frac{\partial}{\partial \phi}\,Y_{l,m}(\theta,\phi)\,.
\end{array}\end{equation}
These operators are not all independent. In fact after the definition of SH in (\ref{5}), 
$-i\,D_\Phi\equiv M$  and, consequently, $-i\,\widetilde D_\Phi\equiv \widetilde M$. It is interesting to realize that eq.~(\ref{6}), valid as an  differential equation defining SH, may be formally rewritten in terms of the above operators, as follows
\begin{equation}\label{38}
\left(\,D^2_\Theta+ \cot\Theta\;D_\Theta+\frac{1}{\sin^2\Theta}\,
D^2_\Phi +L(L+1)\right)\;Y_l^m(\theta,\phi) =0\,,
\end{equation}
\end{widetext}
which can be trivially extended to any {\it finite} linear combination of SH. Some touchy technicalities come after operators  $\cot\Theta$ and $1/{\sin^2\Theta}$, that we shall discuss on Section 4. Just remark that (\ref{38}) is a formal expression from the point of view of operator theory. 

From definitions (\ref{32}), it is clear that $\Theta$ and $\Phi$ have purely absolutely continuous spectrum, which is $[0,\pi]$ in the first case and $[0,2\pi ]$ in the second. They commute, so that they must have a basis of simultaneous eigenvectors, which is often denoted in textbooks as $\{|\theta,\phi\rangle\}$, although  we shall see that this idea should be slightly changed for the benefit of mathematical rigor. These eigenvectors do not belong to the Hilbert space $L^2(S^2,d\Omega)$, although they are properly defined on a larger structure: a rigged Hilbert space. 

On the other hand, the spectrum of both $L$ and $M$ is discrete and degenerate. They also commute and their simultaneous eigenvector basis is $Y_{l,m}(\theta,\phi)$. In this case, these vectors form an orthonormal basis, also called complete orthonormal set, on $L^2(S^2,d\Omega)$, which is the image by $S$ of the orthonormal basis $\{|l,m\rangle\}$ in $\mathcal H$, which is determined by the eigenvectors of the operators $\tilde L$ and $\tilde M$ on $\H$. 

Thus, we are playing with a discrete and a continuous basis, a game that leads us out of the Hilbert space. Again, the situation  will  be saved using RHS. 
%%%%%%%%%%%%%%%%%%%%%%%%%%%%%%%%%%%%%%%%%%%
%%%%%%%%%%%%%%%%%%%%%%%%%%%%%%%%%%%%%%%%%%%

\section{RHS formulation.}

The purpose of this section is to introduce of the above setting within the context of RHS. As mentioned above, RHS is the perfect framework where discrete and continuous bases coexist. In addition, the same RHS serves as a support for a representation of a Lie algebra as continuous operators on it, as well as  for its UEA. Therefore, the connection between discrete and continuous bases and Lie algebras with RHS is well established. 

Now as we have announced, it is the right time to define and use the rigged Hilbert space 
\begin{equation}\label{381} 
\mathfrak G\subset\mathcal H\subset\mathfrak G^\times\,, 
\end{equation}
which will be the arena where discrete and continuous bases coexist and the meaningful operators are well defined and continuous. However, as we have a representation in terms of SH, it would be more convenient 
to start with an equivalent RHS    
\begin{equation}\label{39}
\mathfrak D\subset L^2(S^2,d\Omega)\subset \mathfrak D^\times\,,
\end{equation}
such as  $\mathfrak D$ is a test functions space with functions $f(\theta,\phi)\in L^2(S^2,d\Omega)$, which therefore admit the span \eqref{7}. Thus, having \eqref{7} in mind, we define the test space $\mathfrak D$ as the space of functions $f(\theta,\phi)$ in $L^2(S^2,d\Omega)$ such that
\begin{equation}\label{40}
\big|\big| f(\theta,\phi)\big|\big|_n^2:= \sum_{l,m}
(l+|m|+1)^{2n}\, \big|f_{l,m}\big|^2<\infty\, 
\end{equation}
with $n=0,1,2,\dots\,.$
Obviously, all the finite linear combinations of SH are in $\mathfrak D$, so that $\mathfrak D$ is dense in $L^2(S^2,d\Omega)$. Thus, we have defined a family of norms $ \big|\big|- \big|\big|_n$ on $\mathfrak D$, which gives a topology such that $\mathfrak D$ is a Fr\`echet space (metrizable and complete). Since for $n=0$ we have the Hilbert space norm, the canonical injection from $\mathfrak D$ into $L^2(S^2,d\Omega)$ is continuous. As usual, we denote by $\mathfrak D^\times$ the dual space of $\mathfrak D$.
%%%%%%%%%%%%%%%%%%
%%%%%%%%%%%%%%%%%%

\subsection{Continuity of the generators  of $so(3,2)$ on $\mathfrak D\subset L^2(S^2,d\Omega)\subset \mathfrak D^\times$}
 As we mentioned before, the dynamical algebra of SH, $so(3,2)$, was introduced in \cite{CO}, where
 the explicit expressions as vector fields of the $so(3,2)$ generators are collected.  From our point of view these formal definitions are not interesting and may cause unnecessary difficulties. We shall use an alternative definition of these generators based on their action on SH, that  also appear in \cite{CO},   which is:
 \begin{widetext}
 \begin{equation}\begin{array}{lll}\label{41}
J_\pm\,Y_l^m(\theta,\phi )&:=&\sqrt{(l\mp m)(l\pm m+1)}\,Y_l^{m\pm 1}(\theta,\phi)\,,
\\[0.4cm]
K_+\,Y_l^m(\theta,\phi)&:=& \sqrt{(l+1)^2-m^2}\,Y_{l+1}^m(\theta,\phi)\,,\\[0.35cm]
K_-\,Y_l^m(\theta,\phi)&:=&\sqrt{l^2-m^2}\, Y_{l-1}^m(\theta,\phi)\,,
\\[0.35cm]
R_+\,Y_l^m(\theta,\phi)&:=&\sqrt{(l+m+2)(l+m+1)}\,Y_{l+1}^{m+1}(\theta,\phi)\,,\\[0.35cm]
R_-\,Y_l^m(\theta,\phi)&:=&\sqrt{(l+m)(l+m-1)}\,Y_{l-1}^{m-1}(\theta,\phi)\,,
\\[0.35cm]
S_+\,Y_l^m(\theta,\phi)&:=&\sqrt{(l-m+2)(l-m+1)}\,Y_{l+1}^{m-1}(\theta,\phi)\,,\\[0.35cm]
S_-\,Y_l^m(\theta,\phi)&:=& \sqrt{(l-m)(l-m-1)}\,Y_{l-1}^{m+1}(\theta,\phi)\,.
\end{array}\end{equation}
\end{widetext}
We have to add to generators of  the Cartan subalgebra, since the rank of $so(3,2)$ is 2 and its dimension 10. They are the operators $L$ and $M$ and their  action is given in \eqref{32}.

Since $l$ goes from $0$ to $\infty$, it is clear that the operators (\ref{41}) together with $L$ and $M$  are all unbounded and, therefore, their respective domains are densely defined on the Hilbert space $L^2(S^2,d\O)$, but not on the whole space. 
We can easily prove that all these operators are defined on the whole $\mathfrak D$ and are continuous with the topology on $\mathfrak D$. The proof is simple and it is essentially the same for all  the $so(3,2)$ generators displayed in \eqref{41}. As an example, let us give the proof for $K_+$. For any function $f$ in $\mathfrak D$, we have:
\begin{widetext}\be\begin{array}{l}\label{42}
\ds K_+ f(\theta,\phi)=K_+ \sum_{l,m} 
f_{l,m}\,\sqrt{l+1/2}\,Y_l^m(\theta,\phi)
\;\ds = \sum_{l,m}
 f_{l,m}\,\sqrt{l+1/2}\,\sqrt{(l+1)^2-m^2}\,Y_{l+1}^m(\theta,\phi)\,.
\end{array}\ee\end{widetext}
To show that this vector is in $\mathfrak D$, we have to prove that for any $n=0,1,2,\dots$, it satisfies (\ref{40}), which in this case means that (look at the shift on the index $l$ in (\ref{42})):
\be\label{43}
\sum_{l,m}
 \big|f_{l,m}\big|^2\,((l+1)^2-m^2)\,(l+1+|m|+1)^{2n}\,.
\ee
The following two inequalities are straightforward:
\[\begin{array}{rll}\label{44}
(l+1+|m|+1)^{2n}&\le& 2^{2n}\,(l+|m|+1)^{2n},\\[0.3cm]
(l+1)^2-m^2 &\le & (l+|m|+1)^2\,.
\end{array}\]
Using these inequalities   we ready see that \eqref{43} is bounded by 
\be\label{45}
2^{2n} \sum_{l=0}^\infty \sum_{m=-l}^l  \big|f_{l,m}\big|^2\,(l+1+|m|+1)^{2n+2}\,,
\ee
which obviously converges after (\ref{40}). Therefore,  $K_+\,f\in \mathfrak D$. In order to show the continuity of $K_+$ on $\mathfrak D$, we use the result in (\ref{4}). Thus, let us apply $K_+$ to  an arbitrary $f(\theta,\phi)\in \mathfrak D$. After  (\ref{40}), (\ref{42})  and (\ref{43})   we obtain
 \be \label{46}
   \big|\big|K_+f(\theta,\phi) \big|\big|_n \le 2^n\, \big|\big|f(\theta,\phi) \big|\big|_{n+1}\,,
\ee
which satisfies (\ref{4}) for all $n=0,1,2,\dots$.  Hence, the continuity of $K_+$ on $\mathfrak D$ has been proved. Using the duality formula \eqref{3}, we extend $K_+$ to a weakly continuous operator on $\mathfrak D^\times$. Same properties can be proven for the other generators of $so(3,2)$ displayed in \eqref{41}. 

Now we study the generators of the Cartan subalgebra, $L$ and $M$, that  are also unbounded. 
With a simple operation, we can show that $L$ is well defined as a continuous operator on $\mathfrak D$. Let $f(\theta,\phi)\in\mathfrak D$,  we have that
\begin{equation}\label{56}
L\,f(\theta,\phi)= \sum_{l,m}
 f_{l,m}\,l\,\sqrt{l+1/2}\;Y_l^m(\theta,\phi)\,.
\end{equation}
Considering the norm $ \big|\big|- \big|\big|_n$ given by \eqref{40}, we get  that 
\[
\begin{array}{l}
\big|\big|L\,f(\theta,\phi)\big|\big|_n^2 =\ds \sum_{l=0}^\infty \sum_{m=-l}^l \big|f_{l,m}\big|^2\,l^2 \,(l+|m|+1)^{2n} \\[0.4cm] 
\hskip1.75cm \le \ds\sum_{l=0}^\infty \sum_{m=-l}^l \big|f_{l,m}\big|^2\, (l+|m|+1)^{2n+2}  \\[0.4cm] 
\hskip1.75cm \le \ds \big|\big| f(\theta,\phi)\big|\big|_n^{2n+2}\,,
\end{array}
\]
for $ n=0,1,2,\dots$
The above inequality  shows that $L\mathfrak D\subset \mathfrak D$ as well as the continuity of $L$ on $\mathfrak D$.
The same happens for $M$ since  the proof is similar. Moreover, both operators can be   extended  to   weakly continuous operators on $\mathfrak D^\times$ by means of the duality formula \eqref{3}.

It is straightforward to show that all members of the infinite dimensional algebra spanned by the operators in \eqref{41} are also continuous operators on $\mathfrak D$ as well as on the antidual $\mathfrak D^\times$. 

At this time, it is convenient to define the RHS $\mathfrak G\subset\mathcal H\subset \mathfrak G^\times$. The main tool is the unitary mapping $S$ defined in \eqref{15}. First of all, $\mathfrak G:=S^{-1}\mathfrak D$, hence the topology on $\mathfrak G$ is the transported topology from $\mathfrak D$ by $S$, so that if $|f\r\in\mathfrak G$, the semi-norms are
\begin{equation}\label{47}
\big|\big| |f\r\big|\big|_n^2= \sum_{l,m}
(l+|m|+1)^{2n}\, \big|f_{l,m}\big|^2<\infty\,,
\end{equation}
with $ n=0,1,2,\dots\,.$ 
The topology on $\mathfrak G$ uniquely defines $\mathfrak G^\times$. It is noteworthy that there exists a one-to-one continuous mapping from $\mathfrak G$ onto $\mathfrak D$ with continuous inverse, which is given by an extension of $S$ defined via the following duality formula 
\begin{equation}\label{471}
\langle Sf|SF\rangle=\langle f|F\rangle\,,\qquad |f\r\in\mathfrak G,\; F\in\mathfrak G^\times\,, 
\end{equation}
where we have use the symbol $S$ also to denote its extension. 

On the other hand, if an operator $O$ satisfies $O\mathfrak D\subset \mathfrak D$ with continuity, same property works for $\widetilde O=S^{-1}OS$ on $\mathfrak G$.
%%%%%%%%%%%%%%%%%%%%%%%%
%%%%%%%%%%%%%%%%%%%%%%%%
\subsection{Continuity of other relevant operators}\label{relevantoperators}

Now we shall introduce  some results concerning other  relevant operators for the  discussion on SH. These are $\Phi$, $\Theta$, $-i\,D_\Phi\equiv M$ and $D_\Theta$ whose action on SH is given in \eqref{32}.

The operator $\Phi$  is a bounded operator on $L^2(S,d\Omega)$ as well as $\Theta$.
However, this last operator is relevant  through its functions $\sin\Theta$ and $\cos\Theta$, which are also well defined bounded operators on $L^2(S,d\Omega)$. In addition, both  are continuous operators on $\mathfrak D$ as defined before. 
To prove these properties   we shall use the definition of SH in terms  of the associated Legendre functions \eqref{5} and some properties of these last functions. 

Let us start with $\cos\Theta$.
The first property of the associated Legendre functions used by us  is given by the formula
\begin{widetext}
\[
xP_l^m(x)\\[0.3cm]
=\frac1{2l+1} \left[ (l-m+1) P_{l+1}^m(x)+(l+m)P_{l-1}^m(x)  \right]\,.
\]
 By means of  \eqref{5}, 
we may rewrite this expression  as
\[
\cos\Theta \,Y_l^m(\theta,\phi) 
=\frac1{2l+1} \left[ \sqrt{(l+m+1)(l-m+1)} \;Y_{l+1}^m(\theta,\phi)\right.\\[0.4cm]\ds 
+\left.\sqrt{(l+m)(l-m)}\;Y_{l-1}^m(\theta,\phi)  \right]\,.
\]
Note that the second term in the r.h.s. of this expression  as well as of the previous one vanishes for $l=0$.
For any  $f(\theta,\phi)\in\mathfrak D$, we have
\begin{equation}\begin{array}{lll}\label{50}
\cos\Theta\,f(\theta,\phi)&=&\ds  \left[\sum_{l=0}^\infty \sum_{m=-l}^l f_{l,m} \, \frac1{\sqrt{2}\,\sqrt{2l+1}}\,
\sqrt{(l+m+1)(l-m+1)} \; Y_{l+1}^m(\theta,\phi)   \right.  \\[0.3cm] 
&&\hskip0.25cm \ds\left. + \sum_{l=1}^\infty \sum_{m=-l}^l f_{l,m} \,  \frac1{\sqrt{2}\,\sqrt{2l+1}}\,\sqrt{(l+m)(l-m)} \;Y_{l-1}^m(\theta,\phi)  \right]
 \equiv  g(\theta,\phi)+h(\theta,\phi)\,.
\end{array}\end{equation}
\end{widetext}
In order to show that $\cos\Theta\,f(\theta,\phi)$ is in $\mathfrak D$, provided that 
$f(\theta,\phi)\in \mathfrak D$, we have to prove that both series  in \eqref{50} converge with the topology on  $\mathfrak D$. To do it, it is sufficient to show that both $ \big|\big|f(\theta,\phi) \big|\big|_n$ and  $ \big|\big|g(\theta,\phi) \big|\big|_n$ are finite for all $n=0,1,2,\dots$
If this were the case, we would have
\[\label{51}
\big|\big|\cos\Theta \,f(\theta,\phi)\big|\big|_n\le\big|\big| g(\theta,\phi)\big|\big|_n+
\big|\big| h(\theta,\phi)\big|\big|_n\,.
\]
Thus, we have,
\begin{widetext}
\be\begin{array}{lll}\label{52}
\big|\big| g(\theta,\phi)\big|\big|_n^2  
&=&\ds \sum_{l=0}^\infty \sum_{m=-l}^l \big|f_{l,m}\big|^2 \,\frac1{2 (2l+1)}\,(l-m+1)\,(l+m+1)\,(l+|m|+1)^{2n} 
 \\[0.3cm] 
&\le &\ds  \sum_{l=0}^\infty \sum_{m=-l}^l \big|f_{l,m}\big|^2 \,\frac1{2 (2l+1)}\, (l+|m|+1)^{2n+2}  \\[0.3cm] 
&\le &\ds  \sum_{l=0}^\infty \sum_{m=-l}^l \big|f_{l,m}\big|^2 \, (l+|m|+1)^{2n+2} 
= \ds \big|\big|g(\theta,\phi)\big|\big|^2_{n+1}\,. 
\end{array}\ee 
\end{widetext}
Since  and due to \eqref{40}, the last series in \eqref{52} converges for all values of $n$. This shows the finite character of $\big|\big| g(\theta,\phi)\big|\big|_n$.  
The same result is obtained for $\big|\big| h(\theta,\phi)\big|\big|_n$, so that
\[\label{53}
\big|\big|\cos\Theta \,f(\theta,\phi)\big|\big|_n\le {2}\,\big|\big|f(\theta,\phi)\big|\big|_{n+1}\,.
\]
This last inequality proves at the same time the stability of $\mathfrak D$ under $\cos\Theta$ and the continuity of $\cos\Theta$ on $\mathfrak D$. 

In order to prove that $\sin\Theta\, \mathfrak D \subset \mathfrak D$  and the continuity of $\sin\Theta$ on $\mathfrak S$ we need to use the   following relation of the associated Legendre functions:
\begin{widetext}
\begin{equation}\label{54}
\begin{array}{l}
\sqrt{1-x^2}\,P_l^m(x)
 \quad =\ds \frac{1}{2l+1} [(l-m+1)(l-m+2) \,P_{l+1}^{m-1}(x)-(l+m+1)(l+m)\,P_{l-1}^{m-1}(x)]\,.
\end{array}
\end{equation}
Note that \eqref{54} is well defined as the last term in the r.h.s. vanishes if $l=0$. 
Since $x=\cos\theta$, this equation implies a relation for $\sin\Theta$ similar to (\ref{50}). Then, using essentially the same arguments as above, we prove that $\sin\Theta\,\mathfrak D\subset \mathfrak D$ as well as the continuity of $\sin\Theta$ on $\mathfrak D$.  

Furthermore, if we use the following equation:
\begin{equation}\label{55}
\frac1{\sqrt{1-x^2}}\,P_l^m(x)= -\frac1{2m} [P_{l+1}^{m+1}(x)+(l-m+1)(l-m+2)\,P_{l+1}^{m-1}(x)]\,,
\end{equation}
\end{widetext}
we may prove the same result for $\sin^{-1}\Theta$. This may be a little surprising since $\sin^{-1}\theta$ is not bounded in $[0,2\pi]$. However, note that we are discussing the continuity of operators on a topology different albeit stronger than the Hilbert space topology, on which $\sin^{-1}\Theta$ is not continuous. This also implies the continuity of the operators $\sin^{-2}\Theta$ and $\cot \Theta$ that appear in eq.  \eqref{38}. 

The remaining operators introduced in this Section, $-i\,D_\Phi\equiv M$ and  
$D_\Theta$,  are not bounded on $L^2(S,d\Omega)$. 
In order to study the effect of  $D_\Theta$ on $\mathfrak D$, we have to depart from (\ref{5}) and to take into account some properties of the associated Legendre polynomials. First of all, we note that
\begin{widetext}\[\label{58}
D_\Theta \,Y_l^m(\theta,\phi)=\frac{\partial}{\partial\theta}\, Y_l^m(\theta,\phi)=
C_l^m \,e^{im\phi}\, \frac{d}{d\theta}\, P_l^m(\cos\theta)\,,
\]
where the value of the constant $C_l^m$ has been given in (\ref{5}).  We obtain that
\begin{equation}\begin{array}{lll}\label{59}
\ds\frac{d}{d\theta}\, P_l^m(\cos\theta)&=&\ds\frac{d}{d \theta}\cos\theta\,\frac{d}{d\cos \theta}\,P_l^m(\cos\theta) 
\ds -\sin\theta \,\frac{d}{d\cos \theta}\,P_l^m(\cos\theta) 
\ds- \sqrt{1-x^2}\,\frac{d}{dx}\,P_l^m(x)\,, 
\end{array}\end{equation}
where in the third equality we have made the change of variable $x:=\cos\theta$.
Finally 
 the last terms connects with  a property of the 
associated Legendre polynomials \cite{AS}, so that
\begin{equation}\label{591}
\ds \sqrt{1-x^2}\,\frac{d}{dx}\,P_l^m(x)
  =\ds \frac12\,[(l+m)(l-m+1)\,P_l^{m-1}(x)-P_l^{m+1}(x)]\,.
\end{equation}
 Considering the relations
\[
\frac{(l-m)!}{(l+m)!}= \left\{ \begin{array}{l}
\ds \ds\frac{(l-(m-1))!}{(l+(m-1))!}\,\frac1{(l+m)(l-m+1)}
\\[0.5cm]
\ds\frac{(l-(m+1))!}{(l+(m+1))!}\, (l-m)(l+m+1)
\end{array}\right.\,,
\]
and  taking into account that  
\[
\ds C_l^m= \sqrt{{(l-m)!}/{(l+m)!}}/\sqrt{2 \pi}\,,
\]  eq.~\eqref{591} yields an interesting relation for SH, which is
\[\begin{array}{lll}\label{61}
D_\Theta \,Y_l^m(\theta,\phi) &=& \ds -\frac12\, \left[\frac{1}{\sqrt{(l+m)(l-m+1)}}\,Y_l^{m-1}(\theta,\phi)\right.
-\sqrt{(l-m)(l+m+1)}\,Y_l^{m+1}(\theta,\phi)]\,.
\end{array}\]
Note that for the choices $m=-l$ or $m=l$,  the first or the second term vanishes, respectively, which is consistent with the fact that $-l\le m\le l$. 
Now, let us take an arbitrary $f(\theta,\phi)\in\mathfrak D$. Then,
\begin{equation}\begin{array}{lll}\label{62}
D_\Theta\,f(\theta,\phi) &=&\ds -\frac12 \left[\sum_{l=0}^\infty \sum_{m=-l+1}^l f_{l,m}\,\frac{\sqrt{l+1/2}}{\sqrt{(l+m)(l-m+1)}}\,Y_l^{m-1}(\theta,\phi)\right.  \\[0.4cm] 
&&\hskip1.5cm 
\ds-\left. \sum_{l=0}^\infty \sum_{m=-l}^{l-1}  f_{l,m}\, \sqrt{l+1/2}\, \sqrt{(l-m)(l+m+1)} \,Y_l^{m+1}(\theta,\phi) \right] \,,
\end{array}\end{equation}
that we can rewrite as 
\begin{equation}\label{621}
D_\Theta\,f(\theta,\phi) 
=g(\theta,\phi)+h(\theta,\phi)\,,
\end{equation}
where $g(\theta,\phi)$ and $h(\theta,\phi)$ are, respectively, the first and second of the infinite sums in (\ref{62}). Hence,  
\begin{equation}\label{63}
\big|\big| D_\Theta\,f(\theta,\phi)\big|\big|_n\le \big|\big| g(\theta,\phi)\big|\big|_n
+\big|\big| h(\theta,\phi)\big|\big|_n\,.
\end{equation}
The square of the first norm in the r.h.s. of (\ref{63}) is given by
\begin{eqnarray*}\label{64}
 \big|\big|g(\theta,\phi) \big|\big|_n^2 \le \frac14  \sum_{l=0}^\infty \sum_{m=-l+1}^l \big|f_{l,m}\big|^2\,\frac1{(l+m)(l-m+1)}\,(l+|m-1|+1)^{2n}
\end{eqnarray*}
Since,
\[\begin{array}{rll}\label{65}
\ds \frac1{(l+m)(l-m+1)} \le1\,, 
\qquad\qquad
\ds (l+|m-1|+1)\le (l+|m|+2) \le  (l+|m|+1)^2\,,
\end{array}\]
we conclude that
\[\begin{array}{lll}\label{66}
\big|\big|g(\theta,\phi)\big|\big|_n^2 \le\ds\frac14 \sum_{l=0}^\infty \sum_{m=-l}^l \big|f_{l,m}\big|^2\, 
(l+|m|+1)^{4n} 
= \ds \frac14 \big|\big| f(\theta,\phi)\big|\big|_{2n}^2\,.
\end{array}\]
Similarly, we obtain that
$\ds \big|\big| h(\theta,\phi)\big|\big|_n^2 \le \frac 14\,\big|\big|f(\theta,\phi)\big|\big|^2_{2n+2}\,.
$ In conclusion,
\[\label{68}
\big|\big|D_\Theta\,f(\theta,\phi)\big|\big|_n\le 
\frac 12\left(\big|\big| f(\theta,\phi)\big|\big|_{2n} + \big|\big| f(\theta,\phi)\big|\big|_{2n+2}   \right)\,,
\]
which proves the invariance and continuity of $D_\Theta$ on $\mathfrak S$. 
\end{widetext}

%%%%%%%%%%%%%%%%%%%%%%%%%%
%%%%%%%%%%%%%%%%%%%%%%%%%%
\subsection{The functional $\langle\theta,\phi|$ on $\mathfrak G$.}

This subsection is devoted to define and give some properties of the functional $\langle\theta,\phi|$. As commented before, this object is defined for almost all $\theta$ and $\phi$ with $0\le \theta<\pi$ and $0\le \phi<2\pi$ as a functional on all $|f\r\in\mathfrak G$ as $\langle \theta,\phi|f\rangle:=f(\theta,\phi)=S|f\r$.  We say {\it almost all} because a square integrable function $f(\theta,\phi)\in L^2(S^2,d\Omega)$ is defined save for a set of zero Lebesgue measure. 

Let us fix arbitrary $|f\r\in\mathfrak G$, $\theta$ and $\phi$ and write \eqref{25} as \eqref{18}
where $f_{l,m}$ are the coefficients of the span of $f$ in terms of the discrete basis $\{|l,m\rangle\}$, i.e. $f_{l,m}=\langle l,m|f\rangle$. We want to show that this defines a linear continuous mapping
\begin{equation}\begin{array}{rllll}\label{70}
\langle\theta,\phi| : & \mathfrak G\; &\longmapsto& \; \C
\\[0.3cm]
&|f\r\; &\longmapsto &\;\langle \theta,\phi|f\rangle=f(\theta,\phi) \,,
\end{array}\end{equation}
with $\mathbb C$ the field of complex numbers. The linearity is obvious. The proof for the continuity relies on a version of the inequality \eqref{4} valid for linear mappings.  
It may be stated as follows: let $\Phi\subset\mathcal H\subset \Phi^\times$ a rigged Hilbert space, defined as in Section~\ref{RHSoverview}, and $F$ a {\it linear or anti-linear} mapping from $\Phi$ into the field of complex numbers $\mathbb C$. Then, $F$ is continuous on $\Phi$ if and only if there is a positive number $C>0$ and a finite number of norms defining the topology on $\Phi$, such that for any $|f\r\in\mathfrak G$, we have that
\begin{equation}\label{71}
\big |F(f)\big|\le C\,\{ \big|\big| f\big|\big|_{p_1}+\dots+\big|\big| f\big|\big|_{p_k}\}\,.
\end{equation}
Let us go back to \eqref{18} and let us write it as
\begin{widetext}
\begin{equation}\label{72}
\langle\theta,\phi|f\rangle = \sum_{l=0}^\infty \sum_{m=-l}^l   f_{l,m}\, (l+|m|+1)^{p} \,\left( \frac{\sqrt{l+1/2}}{(l+|m|+1)^{p}} \, Y_l^m(\theta,\phi)\right) \,,
\end{equation}
where $p$ is a natural number such that $p\ge 3$. We may use the Schwarz inequality so as to obtain
\begin{equation}
\label{73}
|\langle\theta,\phi|f\rangle| \le \ds \sqrt{\sum_{l=0}^\infty \sum_{m=-l}^l  \big|f_{l,m}\big|^2\,(l+|m|+1)^{2p}} \;
 \times \;\sqrt{\sum_{l=0}^\infty \sum_{m=-l}^l \frac{l+1/2}{(l+|m|+1)^{2p}}\,
\,\big|Y_l^m(\theta,\phi)\big|^2}\,.
\end{equation}
\end{widetext}
The first series in \eqref{73} converges because $|f\r\in\mathfrak G$ and this series is not more than 
$\big|\big| |f\r\big|\big|_p$  \eqref{47}. In order to see that the second series also converges, we have to take into account that SH have the following upper bound for all values of $\theta$ and $\phi$ \cite{AH}: 
\begin{equation}\label{74}
\big|Y_l^m(\theta,\phi)\big|\le \frac{1}{\sqrt{2\pi}}\,.
\end{equation}
Then and taking into account that $l+|m|+1\ge l+1$, we find the following upper bound for the second series in \eqref{73}:
\be\begin{array}{l}\label{75}
\ds\sum_{l=0}^\infty \sum_{m=-l}^l \frac{l+1/2}{2\pi (l+|m|+1)^{2p}} \\[0.4cm]
\hskip1.5cm \ds \le \sum_{l=0}^\infty \frac{(2l+1)^2}{4\pi (l+1)^{2p}}\,,
\end{array}\ee
where we have performed the sum on $m$ for each value of $l$
because
\[\begin{array}{l}
\ds\sum_{m=-l}^l \frac{1}{ (l+|m|+1)^{2p}} \le\sum_{m=-l}^l \frac{1}{ (l+1)^{2p}} \\[0.35cm] 
\hskip1.5cm\le\ds
(2l+1)\, \frac{1}{ (l+1)^{2p}} \,.
\end{array}\]
 For $p\ge 2$, the r.h.s. of \eqref{75}, obviously converges. Then, the use of the Schwarz inequality makes sense. Furthermore, if we call $C$ to the value of the second square root in \eqref{73}, we conclude that for any $|f\r\in\mathfrak G$ and {\it any  fixed} $p=2,3,4,\dots$, one has
\begin{equation}\label{76}
\big|\langle\theta,\phi|f\rangle\big |\le C \, \big|\big| f\big|\big|_p\,.
\end{equation}
Following \eqref{71}, this proves the continuity of the linear functional $\langle \theta,\phi|$ on $\mathfrak G$ for fixed $\theta$, $\phi$. 

Since $\langle f|\theta,\phi\rangle=\langle \theta,\phi|f\rangle^*$, it follows that $|\theta,\phi\rangle$ is a continuous anti-linear functional on $\mathfrak G$ and, therefore, $|\theta,\phi\rangle \in\mathfrak G^\times$.
%%%%%%%%%%%%%%%%
%%%%%%%%%%%%%%%%
\subsection{Properties of $|\theta,\phi\rangle\in\mathfrak G^\times$.}

Let us remember that  $S:\mathcal H\to  L^2(S^2,d\Omega)$ is a unitary mapping \eqref{15} with the property that  $S:\mathfrak G \to \mathfrak D$ continuously and with continuous inverse, and  that the duality formula \eqref{471} $\langle F|f\rangle= \langle SF|Sf\rangle$ defines,  for all $|f\r\in\mathfrak G$ and $F\in\mathfrak G^\times$, an extension of $S$ to the duals so that $S:\mathfrak G^\times \to \mathfrak D^\times$ is continuous with continuous inverse. Therefore, $S|\theta,\phi\r$ exists and is in $\mathfrak D^\times$. Then,  we define $\widehat{\cos\Theta}:=S^{-1}\,\cos\Theta\,S$, which by construction is a continuous operator from $\mathfrak G$ into itself.  Thus,
\begin{widetext}
\be\begin{array}{lll}\label{77}
\cos\Theta\,S|f\rangle=\cos\Theta\,f(\theta,\phi)=\cos\theta\,f(\theta,\phi)
=g(\theta,\phi)\in\mathfrak D\,.
\end{array}\ee
Then,
\[\begin{array}{lll}\label{78}
\langle\theta,\phi| \widehat{\cos\Theta}|f\rangle &=& \langle\theta,\phi| S^{-1}{\cos\Theta}S|f\rangle
=
\langle\theta,\phi| S^{-1}\,g(\theta,\phi)\rangle
= g(\theta,\phi)
=
\cos\theta\,f(\theta,\phi) 
= \cos\theta \langle\theta,\phi|f\rangle\,.
\end{array}\] 
Since any continuous operator on $\mathfrak G$ may be extended to a weakly continuous operator on $\mathfrak G^\times$ using the duality formula, then, we conclude that
\begin{equation}\label{79}
\widehat{\cos\Theta}|\theta,\phi\rangle=\cos\theta\,|\theta,\phi\rangle\,,
\end{equation}
for any $0\le\theta<\pi$ and arbitrary $0\le\phi<2\pi$. 

A similar result for the coordinate $\phi$ may be obtained. First of all, let us recall that, in agreement with \eqref{5}, the  product law for SH is \cite{BS}:
\begin{equation}\label{80}
Y_{l_1}^{m_1}(\theta,\phi)\,Y_{l_2}^{m_2} (\theta,\phi) 
=\frac{1}{\sqrt{2\pi}}\,\sum_{L,M}\langle l_1\,0\,l_2\,0|L\,0\rangle\,\langle l_1\,m_1\,l_2\,m_2|L\,M\rangle\,Y_L^M(\theta,\phi)\,,
\end{equation}
\end{widetext}
with $M=m_1+m_2$ and $L$  such that $|l_1-l_2|\le L\le l_1+l_2$ and compatible with $M$, i.e. 
$|M|\le L$.  In particular, we have:
\be\begin{array}{l}\label{81}
Y_1^1(\theta,\phi)\,Y_l^m(\theta,\phi)\\[0.3cm]
\ds \hskip1,5cm =\frac{1}{\sqrt{2\pi}}\,\sum_{L,M}\,c_{CG} \,Y_L^M(\theta,\phi)\,,
\end{array}\ee
where $c_{CG}$ are the above Clebsch-Gordan coefficients for the different values of $l$ and $m$, i.e. 
\[
c_{CG}\equiv c_{CG}(l,m;L,M)=\langle 1\,0\,l\,0|L\,0\rangle\,\langle 1\,1\,l\,m|L\,M\rangle
\]
Note that  
$L$ only takes the values $|l-1|,l$ and $l+1$. Moreover the fact that $\langle 1\,0\,l\,0|l\,0\rangle=0$ for all $l$ implies that  for  $l\neq 0$ in the expression \eqref{81} can only  appear  terms related with $L=|l-1|$ and with $L=l+1$.
For $l-1$ and $l+1$ we find the same value  $L=l$ unless $L=0$ that appears only one time.

Let us consider an arbitrary $f(\theta,\phi)\in\mathfrak D$. Hence it can be written in the basis of  SH in the form \eqref{7}. Then we can write using \eqref{81} that 
\begin{widetext}
\[
Y_1^1(\theta,\phi)\,f(\theta,\phi)=\frac{1}{\sqrt{2\pi}}\,\sum_{l=0}^\infty \sum_{m=-l}^l f_{l,m}\,\sqrt{l+1/2}\,\sum_{L,M} \, c_{CG} \,Y_L^M(\theta,\phi)\,
\]
An estimation   for $\big|\big|Y_1^1(\theta,\phi)\,f(\theta,\phi)\big|\big|^2_p$ gives
\[
\begin{array}{lll}\label{82}
\big|\big|Y_1^1(\theta,\phi)\,f(\theta,\phi)\big|\big|_p^2 \le\ds
\sum_{l,m}
(2l+1)^2\, (l+|m|+1)^{2p}\,|f_{l,m}|^2 
\le \ds  \sum_{l,m}
(l+|m|+1)^{2(p+1)}\,|f_{l,m}|^2 =\big|\big|f(\theta,\phi)\big|\big|^2_{p+1}\,,
\end{array}\]
\end{widetext}
where we have consider that $2 l+1\leq 2(l+|m|+1)$, relation \eqref{74} and that the Clebsch-Gordan coefficients must have a modulus equal or smaller to one. 
Thus, the multiplication by $Y_1^1(\theta,\phi)$ is a linear continuous operator on $\mathfrak D$. Observe that from the explicit expression for  $Y_1^1(\theta,\phi)$  we have that
\[\begin{array}{lll}\label{83}
Y_1^1(\theta,\phi)\,f(\theta,\phi) &=&\ds  C \,\sin\theta\,e^{i\phi}\, f(\theta,\phi) \\[0.3cm]
&=&\ds C\,\sin\Theta\,e^{i\Phi}\,f(\theta,\phi)\,,
\end{array}\]
where $C$ is the corresponding  normalization constant. Thus, the operator $\sin\Theta\,e^{i\Phi}$ is continuous on $\mathfrak D$. In Section~\ref{relevantoperators} we have proven that $\sin^{-1}\Theta$ is also continuous on $\mathfrak D$, then, so is $e^{i\Phi}$.  Therefore, $\widehat {e^{i\Phi}}:=S^{-1}\,e^{i\Phi}\,S$ is continuous on $\mathfrak G$. By an analogous reasoning to the previous situation, we may extend this operator to a weakly continuous linear mapping on $\mathfrak G^\times$, so that
\[
\widehat {e^{i\Phi}} \,|\theta,\phi\rangle=e^{i\phi}\,|\theta,\phi\rangle\,.
\]
This is the last formula we wanted to show in this section, concerning the properties of the functional $ |\theta,\phi\rangle$.

%%%%%%%%%%%%%%%%%%%%%%%%%%%%%%%%%%%%%%%%%%%%%%%
%%%%%%%%%%%%%%%%%%%%%%%%%%%%%%%%%%%%%%%%%%%%%%%

\section{Concluding remarks}

Discrete and continuous bases are quite often used in quantum mechanics. While the former may be identified with complete orthonormal sets in a separable infinite dimensional Hilbert space, the second does not play any role on a Hilbert space. It is in the context of rigged Hilbert spaces, where both types of bases acquire full meaning. The continuous bases are a set of functionals over a space of test vectors. Some formal relations between both types of bases are given.

From the point of view of the algebra $so(3,2)$ some of its members are bounded from the point of view of Hilbert spaces and some others do not. In our rigged Hilbert space, unbounded operators in $so(3,2)$ becomes continuous. However, this is not the case for those which are bounded, which do not even preserve the spaces of test vectors. This is in principle unfortunate, since we expect that vectors in the continuos basis be (generalized) eigenvectors of these operators. Fortunately, this disease has a cure, as we may construct some functions of these operators with the required properties.

%%%%%%%%%%%%%%%%%%%%%%%%%%%%%%%%%%%%%%%%%%%%%%
\section*{Acknowledgments}
This work was partially supported by the Ministerio de Econom\'ia y Competitividad of Spain (Project MTM2014-57129-C2-1-P with EU-FEDER support) and the Junta de Castilla y Le\'on (Project VA057U16).

\end{document}